\documentclass[preprint,12pt]{elsarticle}

\usepackage{amssymb}
\usepackage{xurl}

\usepackage[hidelinks]{hyperref}
\usepackage{graphicx}
\usepackage{amsmath}
\usepackage{cleveref}
\usepackage{booktabs}
\usepackage{siunitx}
\usepackage{svg}
\usepackage{multirow}
\usepackage{geometry}
\journal{International Journal of Multiphase Flow}

\begin{document}

\begin{frontmatter}



\title{Experimental and Numerical Study of Microcavity Filling Regimes for Lab-on-a-Chip Applications}

\author[inst1,inst3]{Luise Nagel}
\author[inst1]{Anja Lippert\corref{cor1}}
\ead{Anja.Lippert@bosch.com}
\author[inst1]{Ronny Leonhardt}
\author[inst1]{Tobias Tolle}
\author[inst2,inst3]{Huijie Zhang}
\author[inst3]{Tomislav Mari\'c\corref{cor1}}
\ead{maric@mma.tu-darmstadt.de}

\cortext[cor1]{Corresponding author}

\affiliation[inst1]{organization={Corporate Research, Robert Bosch GmbH},
            addressline={Robert-Bosch-Campus 1}, 
            city={Renningen},
            postcode={71272}, 
            country={Germany}}
            
\affiliation[inst2]{organization={Mobility Electronics, Robert Bosch GmbH},
            addressline={Markwiesenstrasse 46}, 
            city={Reutlingen},
            postcode={72770}, 
            country={Germany}}

\affiliation[inst3]{organization={Mathematical Modeling and Analysis, Technical University of Darmstadt},
            addressline={Peter-Grünberg-Str. 10}, 
            city={Darmstadt},
            postcode={64287}, 
            country={Germany}}
            
\begin{abstract}
\small
The efficient and voidless filling of microcavities is of great importance for Lab-on-a-Chip applications.
However, predicting whether microcavities will be filled or not under different circumstances is still difficult due to the local flow effects dominated by surface tension. 
In this work, a close-up study of the microcavity filling process is presented, shedding light on the mechanisms of the filling process using experimental insights accompanied by 3D numerical simulations.
The movement of a fluid interface over a microcavity array is investigated optically under consideration of different fluids, capillary numbers, and cavity depths, revealing a regime map of different filling states.
Moreover, the transient interface progression over the cavities is analyzed with attention to small-scale effects such as pinning.
Besides the visual analysis of the image series, quantitative data of the dynamic contact angle and the interface progression is derived using an automated evaluation workflow. 
In addition to the experiments, 3D Volume-of-Fluid simulations are employed to further investigate the interface shape.
It is shown that the simulations can not only predict the filling states in most cases, but also the transient movement and shape of the interface.
The data and code associated with this work are publicly available at Bosch Research GitHub~\cite{github_microcavityFilling} and at the TUDatalib data repository~\cite{tudatalib}.
\textcolor{red}{
\\ This is an arXiv preprint. Please refer to and cite the published article:
L. Nagel, A. Lippert, R. Leonhardt, T. Tolle, H. Zhang, T. Mari\'c. Experimental and numerical study of microcavity filling regimes for Lab-on-a-Chip applications. International Journal of Multiphase Flow 188 (July 2025), p. 105208. DOI:}
\textcolor{blue} {\href{https://doi.org/10.1016/j.ijmultiphaseflow.2025.105208}{10.1016/j.ijmultiphaseflow.2025.105208}}

\end{abstract}



\begin{keyword}
\small
Lab-on-a-Chip \sep Volume-of-Fluid \sep Interfacial flow \sep Microfluidics \sep Contact angle
\end{keyword}

\end{frontmatter}

\section{Introduction}
\label{sec:introduction}
Lab-on-a-Chip (LoC) systems are an important area of research in healthcare technology.
Providing fast and automated molecular diagnostic workflows through miniaturization, LoC technology makes it possible to perform personalized medical analysis at the point of care without the need of large laboratories or specially trained personnel.
LoC systems exploit microfluidic processes for the transport of samples and reagents~\cite{pandey_microfluidics_2018, niculescu_fabrication_2021}.
A prominent example is the filling of microcavities with sample liquids.
Microcavity arrays can be employed for the application of multiplex quantitative Polymerase Chain Reaction (qPCR)~\cite{morrison_nanoliter_2006, ahrberg_microwell_2019, podbiel_fusing_2020}.
The aliquotation of the sample fluid into separate isolated nanoliter volumes allows for highly parallelized sample analysis.
Further applications of microcavity arrays are the trapping and separation of cells from fluid samples~\cite{rettig_large-scale_2005, yin_microfluidics-based_2019}.
A prerequisite to the efficiency of these complex fluidic processes is the understanding of the underlying multiphase flow mechanisms during the filling process, as recently discussed in~\cite{podbiel_analytical_2020, nagel_stabilizing_2024}.
\par 
Experimental work on this topic typically features the large-scale filling of arrays consisting of hundreds of cavities, as investigated in~\cite{liu_rapid_2009,cui_facile_2021}.
The local transient flow behavior, including the contact line pinning at sharp edges, and local deformation of the interface are not considered in these large-scale experiments.
In~\cite{padmanabhan_enhanced_2020}, the filling efficiency of microcavities is investigated under consideration of different flow velocities, cavity geometries, and contact angles, focusing on the steady state after the liquid has passed the cavities.
In addition to experiments, simulation results are shown in~\cite{padmanabhan_enhanced_2020}, indicating that simulations can complement experimental data by giving 3D insight in addition to the top-down 2D view acquired from experiments.
However, the transient processes causing the different filling efficiencies are not investigated further.
A closer look at the transient filling behavior is given for very specific geometries, such as in~\cite{sposito_staggered_2017, lin_scalable_2020, quan_robust_2024, cohen_self-digitization_2010}.
However, these insights are not generalizable to standard topologies such as cylindrical cavities, which are the most common shapes in LoC applications.
\par In recent work~\cite{nagel_stabilizing_2024}, we consider two scenarios of flow over cylindrical microcavities using 3D Volume-of-Fluid (VoF) simulations.
To stabilize parasitic currents, we employ and validate an artificial viscosity model in~\cite{nagel_stabilizing_2024}.
In the present work, we extend~\cite{nagel_stabilizing_2024} with a comprehensive experimental and numerical study of the cavity filling process.
\par In order for simulations to represent two-phase flow including wetting effects, a meaningful choice of the contact angle model is of high importance.
The contact angle is the angle formed between the fluid interface and a solid surface.
In the absence of external forces, an equilibrium contact angle $\theta_{eq}$ can be reached, which is rare in technical applications.
In a system with a moving interface, the dynamic contact angle depends on the contact line speed as well as local effects such as roughness and chemical structure, as is well known for decades~\cite{de_gennes_wetting_1985}.
However, although many modeling approaches exist to describe the contact angle dynamics, as recently reviewed in~\cite{mohammad_karim_physics_2022}, currently available models are still not predictive for a given material combination due to the combination of effects ranging from the molecular to the macroscopic scale.
Furthermore, the implementation of a contact angle model into the finite volume method is not trivial and gives rise to a large number of approaches, as reviewed in~\cite{afkhami_challenges_2022}.
The most straightforward approach is to measure an apparent contact angle in experiments for a given velocity and providing it as a boundary condition in the simulations.
This method, however, requires a high quality and reproducibility of the experimental measurements, which is not straightforward in dynamic cases, as recently discussed in~\cite{zhang_experimental_2024}.
\par This work presents a thorough close-up investigation of the microcavity filling process through experimental and numerical investigation.
The main objective is to identify different filling regimes and the corresponding movement of the fluid interface.
The question is: under which circumstances are cavities filled, not filled, or partially filled?
To this end, experiments are performed to visualize the interface shape and movement in a setting of different fluids, velocities and geometries relevant to LoC applications.
The experimental data is post-processed in an automated and thus reproducible workflow, yielding quantitative data for the interface movement and the dynamic contact angle.
To complement the 2D experimental data, 3D VoF simulations are performed.
They are calibrated using the experimentally acquired contact angles.
The simulations are compared to the experiments with regard to the filling regimes and interface progression, showing very good agreement and providing insight into the interface shape beyond the experimental view.
The complete dataset and the post-processing scripts are made publicly available at Bosch Research GitHub~\cite{github_microcavityFilling} and at the TUDatalib data repository~\cite{tudatalib}.

\section{Materials and Methods}
\label{sec:materials_methods}
%
\subsection{Experimental Methods}
\subsubsection{Materials and Setup}
\label{subsubsec:exp_materials}
\begin{figure*}
    \centering
    \includegraphics[width=\textwidth]{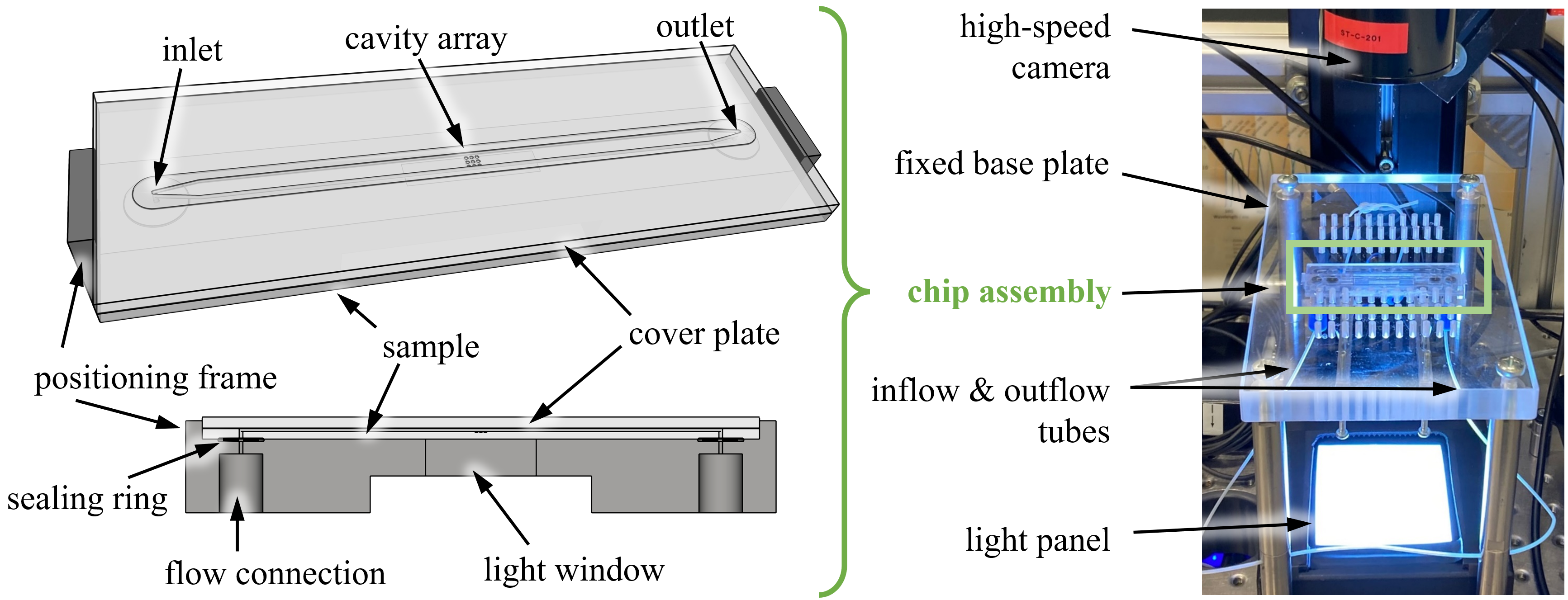}
    \caption{Setup of the sample chip and the measurement system.}
    \label{fig:exp_setup}
\end{figure*}
\begin{figure}
    \centering
    \fontsize{9}{10}\selectfont
    \includegraphics[width=\textwidth]{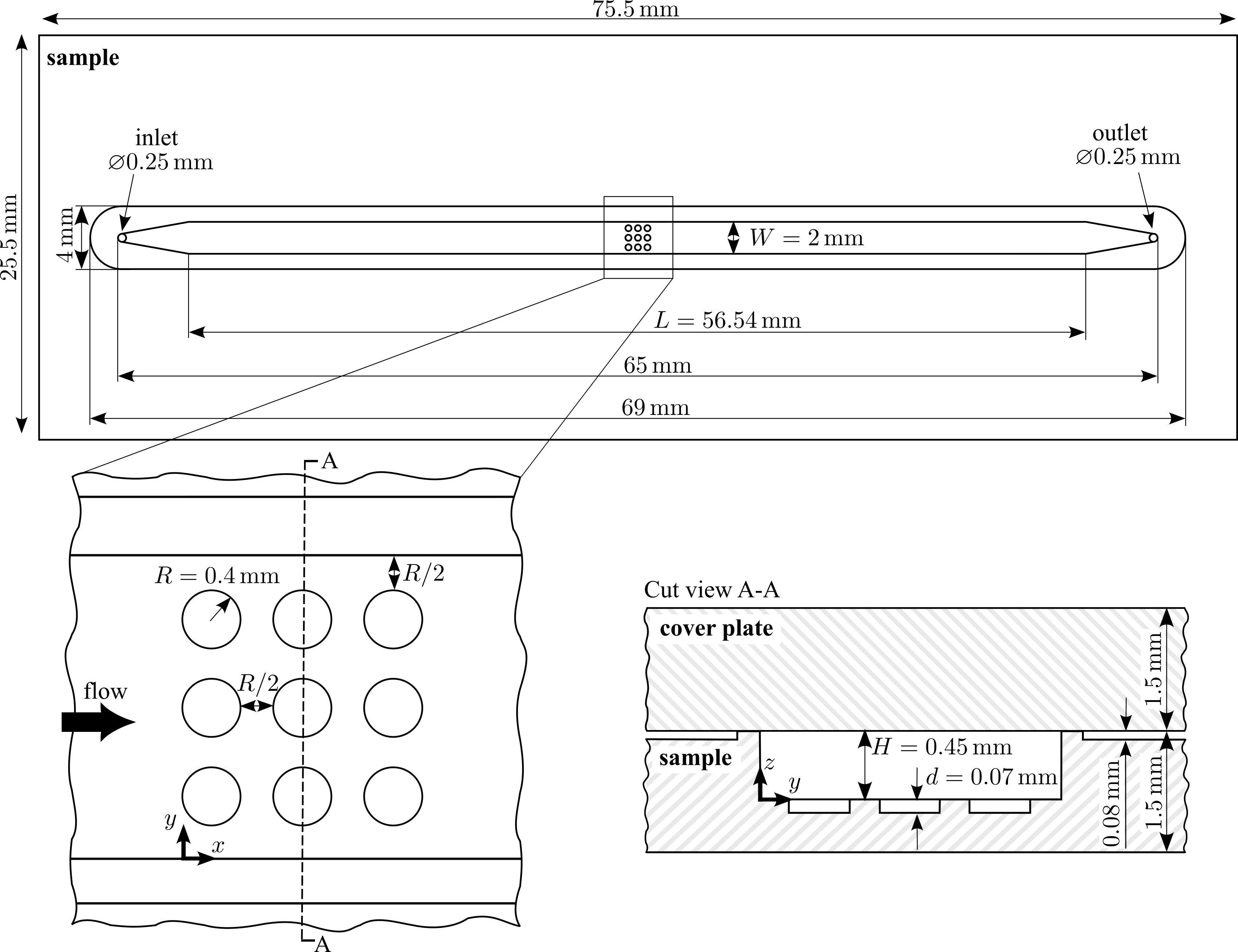}
    \caption{Geometry and measurements of the sample used for the experiments. In the cut view (bottom right), the cover plate is also shown. In selected experiments, the cavity depth is increased to $d=\SI{0.14}{mm}$. }
    \label{fig:exp_geometry}
\end{figure}
The experimental setup is shown in~\cref{fig:exp_setup} and explained in detail in~\cite{nagel_stabilizing_2024}, a summary is given in the following.
The flow over microcavities is visualized using top-down camera recordings.
As test samples, polymethyl-methacrylate (PMMA) microscopy slides (Microfluidic ChipShop\textcolor{black}{~\cite{chipShop}}) are customized with open channels and cavities by micromilling.
The resulting geometry and measurements are given in~\cref{fig:exp_geometry}. 
The cavity depth is $d=\SI{0.07}{mm}$.
For the regime map study in~\cref{subsubsec:regime_map}, a second variant with deeper cavities is considered, $d=\SI{0.14}{mm}$.
As a result of the micromilling, the sample roughness is in the scale of $\leq \SI{5e-6}{m}$.
A smooth, transparent cover lid (Zeonex, Microfuidic ChipShop\textcolor{black}{~\cite{chipShop}}) is used to close the channels. 
The flow is controlled by a syringe pump, and a high-speed camera is used to record the flow over the microcavity array.
The images are resolved with a pixel size of three micrometers.
Three working fluids are employed: water, a $\SI{0.1}{\percent}$ Tween$^{\text{\tiny{\textregistered}}}$-80~aqueous solution (Tween$^{\text{\tiny{\textregistered}}}$-80, Croda International plc), and Novec${\text{\texttrademark}} \, 7500$ (3M${\text{\texttrademark}}$ Novec${\text{\texttrademark}} \, 7500$ Engineered fluid).
The fluid properties are given in~tab.~\ref{tab:fluid_params}.
The Tween$^{\text{\tiny{\textregistered}}}$-80~aqueous solution, hereafter referred to as Tween, is typically used as a buffer solution in LoC applications and differs significantly from water in the surface tension parameter~\cite{szymczyk_aggregation_2016, rehman_surface_2017, szymczyk_effect_2018}.
Novec${\text{\texttrademark}} \, 7500$, hereafter called Novec, which can be used as a sealing fluid in LoC microcavity arrays, exhibits very low surface tension.
Furthermore, it shows perfect wetting on PMMA and Zeonex surfaces ($\theta_{eq}=0$), whereas both other liquids are partially wetting ($\theta_{eq}>0$).
Four different volumetric flow rates are considered.
The resulting mean inflow velocities $U$ are given in~tab.~\ref{tab:ca_exp}.
Additionally, the corresponding capillary numbers are provided in~tab.~\ref{tab:ca_exp}.
The capillary number is defined as
\begin{equation}
    \label{eq:capillaryNumber}
     Ca = \dfrac{ \mu U } {\sigma},
\end{equation}
where $\mu$ is the dynamic viscosity of the liquid phase and $\sigma$ is the surface tension coefficient.
The capillary numbers are in the range of $Ca= \SI{e-6}{}... \SI{e-3}{}$, which is the regime relevant to LoC applications.
The  investigation of lower capillary numbers and static contact angles in this setting is not feasible with satisfactory accuracy, since the interface movement appears unstable and very sensitive to small roughness edges in the geometry.
\begin{table*}[]
\footnotesize
\centering
\caption{Fluid parameters and surface tension coefficients in fluid-air at $\SI{20}{\degree C}$ temperature. Surface tension coefficient of Tween$^{\text{\tiny{\textregistered}}}$-80~solution - air was measured in-house.}
\label{tab:fluid_params}
\begin{tabular}{@{}llll@{}}
    \toprule
    Fluid                                                & Density $\rho \ (\SI{}{kg/m^3}) $ & Dyn. viscosity $\mu \ (\SI{}{m\pascal \second})$ & 
    Surface tension $\sigma \ (\SI{}{mN/m}) $\\ 
    \midrule

    Air \cite{Wagner2013}                                  & $1.19$    & $\SI{0.0182}{}$            & -                          \\
    Water\cite{Lemmon2022, Wagner2013,IAPWS2014}                                                & $998.2$   & $\SI{0.9982e-3}{}$         & $0.07274$                   \\
    Tween$^{\text{\tiny{\textregistered}}}$-80~solution \cite{szymczyk_aggregation_2016}  & $998.2$   & $\SI{0.9982e-3}{}$         & $0.03500$                   \\
    Novec-7500 \cite{Novec2022}                                          & $1614.0$  & $\SI{1.2428e-3}{}$         & $0.01620$                   \\ 
    \bottomrule
\end{tabular}
\end{table*}
\begin{table*}[]
\footnotesize
\centering
    \caption{Capillary numbers corresponding to the given velocities for each experimental case.}
    \label{tab:ca_exp}
    \begin{tabular}{@{}lllll@{}}
        \toprule
        Velocity U (m/s)  &  $Ca$ Water            & $Ca$ Tween             & $Ca$ Novec      \\ \midrule
        $\SI{e-4}{}$   & $\SI{1.37e-6}{}$  & $\SI{2.85e-6}{}$  &       \\
        $\SI{e-3}{}$   & $\SI{1.37e-5}{}$  & $\SI{2.85e-5}{}$  &        \\
        $\SI{e-2}{}$   & $\SI{1.37e-4}{}$  & $\SI{2.85e-4}{}$  & $\SI{7.67e-4}{}$  \\
        $\SI{e-1}{}$   &                   &                   & $\SI{7.67e-3}{}$  \\ \bottomrule
    \end{tabular}
\end{table*}
\subsubsection{Data Processing}
\label{subsec:exp_data_processing}
%
\begin{figure*}
    \centering
    \includegraphics[width=\textwidth]{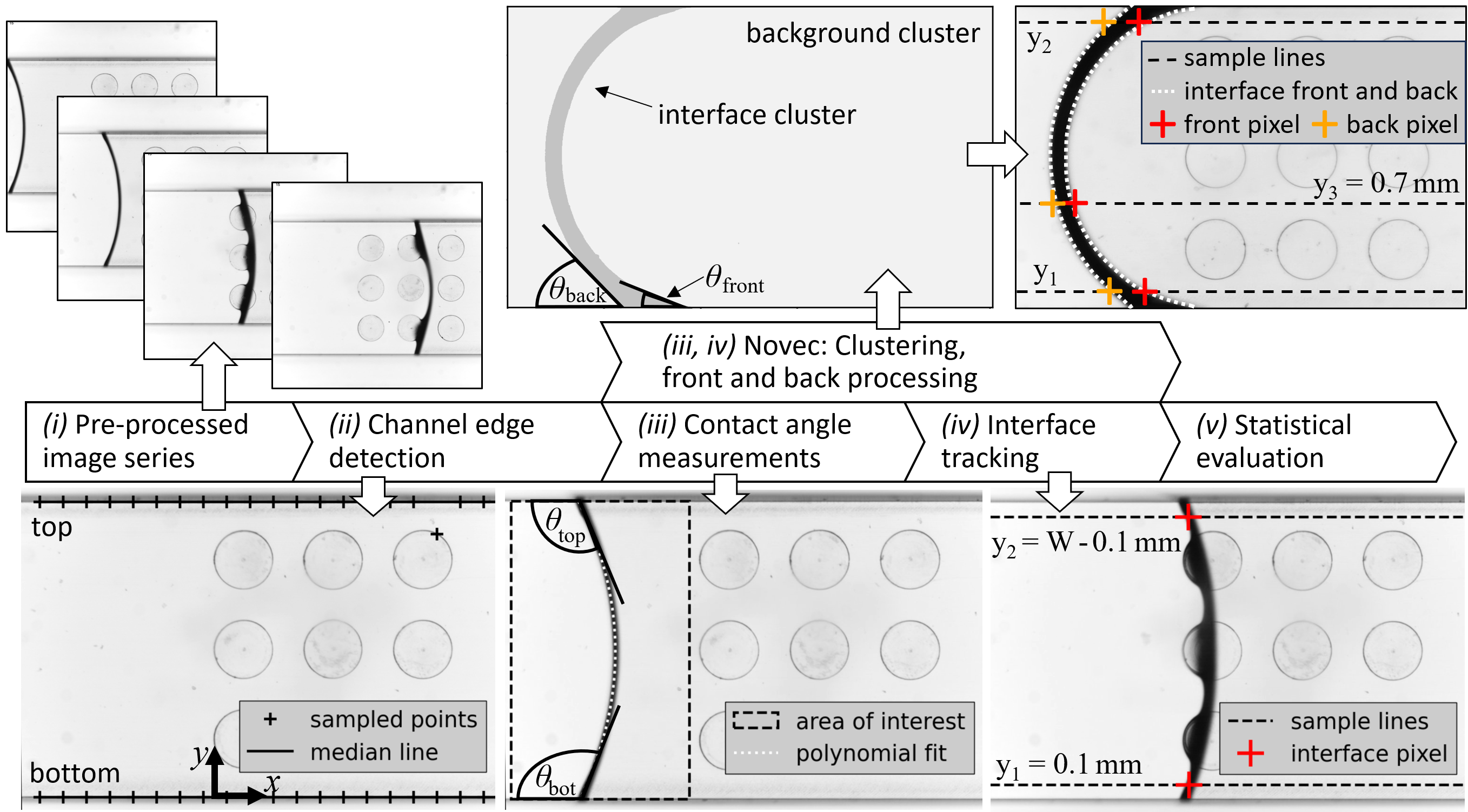}
    \caption{Workflow of the experimental data processing.}
    \label{fig:exp_data_workflow}
\end{figure*}
Based on the recorded time series of the interface movement, three sets of information are derived: First, the filling regimes of the cavities, second, the dynamic contact angles upstream of the cavity array, and third, the quantitative interface progression over time.
For the first task, a manual evaluation is sufficient, since only a qualitative analysis is necessary, see~\cref{subsubsec:exp_filling_regimes}.
The workflow for the contact angle measurement and interface tracking is automated using Python.
The automation provides a basis for standardized and reproducible measurements.
It is presented in~\cref{fig:exp_data_workflow} and published together with the complete dataset at~\cite{github_microcavityFilling, tudatalib}.
The processing steps are shortly described hereafter.
\subparagraph{(i) Pre-processed image series}
The only manual steps that need to be taken are the rotation of the images to ensure the horizontal alignment of the channel edges, and the definition of the x-range of the area of interest for the contact angle detection (step \textit{(iii)}). 
The cavity array is excluded from the contact angle measurements because influences of the local topology on the contact angles are expected in that area.
These steps are carried out using the software ImageJ~\cite{schindelin_fiji_2012}.
\subparagraph{(ii) Channel edge detection}
The horizontal channel edges, which appear dark in the images, are detected using the Python library OpenCV~\cite{opencv_library} and a sampling method.
The lowest grayscale values (darkest pixels) across multiple vertical lines are sampled.
The line positions are chosen such that the channel edges are known to be within the sampling range of the lines.
The median of the $y$-positions of the sampled darkest pixels is chosen as the position of the channel edge.
The choice of the median is, provided many sample lines, found to be very robust against outliers. 
\textcolor{black}{The sampled points and the median lines are shown in the bottom left subfigure of~\cref{fig:exp_data_workflow}.
The channel edges are identified as \textit{bottom} at $y=0$ and \textit{top} at $y=W$, where $W$ is the channel width.}
\subparagraph{(iii) Contact angle measurement}
The contact angles are measured in a global approach which is schematically shown in the center bottom subfigure in~\cref{fig:exp_data_workflow}.
The whole interface, which appears dark in the images, is approximated with a polynomial fit.
A polynomial degree of three is found to be robust and to approximate the interface sufficiently well.
The fitting is preceded by a simple filtering step to remove outliers and is performed with the Random Sample Consensus (RANSAC) method~\cite{fischler_random_1987} available in the scikit-learn library~\cite{scikit-learn}, which includes an additional filter for outliers.
Once the polynomial representation of the interface is acquired, the contact angles at the channel bottom and top are easily computed from the slope at the edges of the interface.
\subparagraph{(iv) Interface tracking}
To quantify the interface progression, a point tracking method is used.
At each time step, the darkest pixel across two horizontal lines is detected.
The lines are located at $y_1=\SI{0.1}{mm}$ and $y_2=W-\SI{0.1}{mm}$.
As shown in the bottom right image in~\cref{fig:exp_data_workflow}, these positions correspond to the center between the channel walls and the respective adjacent cavity rows.
The positions are chosen as a compromise between closeness to the cavities and robustness of the algorithm since the tracking of the interface directly at a cavity is prone to sampling errors due to cavity edges appearing as dark regions in the image.
\subparagraph{(v) Statistical evaluation}
The measured contact angles are collected for all time steps of each experiment.
This large dataset is evaluated in box plots that show the measurement distributions. 
The results are discussed in~\cref{subsec:results_ca_exp}.
Furthermore, the accuracy of the fitting algorithm in step~\textit{(iii)} is evaluated via the distance between the sample point inliers and the resulting fit curve, as shown in~\ref{app:rms}.
%
\subsubsection{Additional Post-processing  for Interface with Low Contact Angle}\label{subsec:porproc_novec}
In experiments with Novec, the low contact angles result in increased interface thickness due to 3D effects.
Consequently, the default interface detection and tracking algorithms~\textit{(iii, iv)} yield high uncertainties, which require alternative post-processing steps.
First, the interface is extracted using the clustering algorithm KMeans~\cite{scikit-learn} which separates the image into two clusters: the interface and the background.
This is shown in~\cref{fig:exp_data_workflow} in the top center subfigure.
From the interface cluster, the front and back are identified.
As seen in the image, the apparent contact angle at the front and back of the interface are not necessarily equal.
Therefore, they are measured and stored separately as $\theta_{front}$ and $\theta_{back}$, respectively.
To account for the elongated front, only the wall-adjacent pixels are used for the polynomial fit.
In the interface tracking step, an additional horizontal line between the cavities is considered ($y_3=\SI{0.7}{mm}$) to account for the strongly curved interface.
This is shown in the top right image of~\cref{fig:exp_data_workflow}.
The front and back of the interface are tracked separately, as shown in the image.
Further details and all fitting parameters are documented in the code published at~\cite{github_microcavityFilling, tudatalib}.
\subsection{Simulation Methods}
\subsubsection{Mathematical and Numerical Modeling}
The simulations are performed with the plicRDF-isoAdvection geometrical unstructured VoF method~\cite{hirt_volume_1981, roenby_computational_2016, scheufler_accurate_2019}.
The reader is directed to a review of unstructured geometrical VoF methods for more details~\cite{maric_unstructured_2020}.
As numerical framework, the open-source finite volume library OpenFOAM-v2212~\cite{OpenFOAMcode} is used.
Specifically, the employed solver is interFlow, part of the OpenFOAM-based TwoPhaseFlow framework introduced in~\cite{scheufler_twophaseflow_2023, TwoPhaseFlowCode}.
This solver was recently validated for wetting cases in~\cite{asghar2023numerical}, showing good results.
To increase the simulation robustness, an artificial interface viscosity and a wisp filtering method are used, which are discussed in detail in~\cite{nagel_stabilizing_2024}.
The chosen artificial viscosity is $\mu_\Sigma = \mu$, where $\mu$ signifies the dynamic viscosity of the liquid phase, and the wisp tolerance is $\varepsilon_w=\SI{e-5}{}$.
These methods were validated for different hydrodynamic and wetting cases for which the reader is referred to~\cite{nagel_stabilizing_2024}.
\subsubsection{Setup}
The case setup is made publicly available at~\cite{github_microcavityFilling, tudatalib} to ensure reproducability in future studies.
The geometry used in the simulations is identical to the experimental geometry given in~\cref{fig:exp_geometry}.
The only difference are the inflow length upstream of the cavity array and the outflow length downstream of the array. 
Instead of simulating the entire channel length $L$, the inflow length~$L_{\mathrm{inflow}}$ upstream of the cavity array and the outflow length~$L_{\mathrm{outflow}}$ downstream of the cavity array are adapted to the case parameters such that (1) the interface shape can stabilize before hitting the cavities, and (2) the interface does not reach the outlet before the whole cavity array is wetted.
The lengths are given for each case in tab.~\ref{tab:sim_params} along with the contact angles used for each simulation, which are further explained below.
An $xz-$symmetry plane is employed to reduce the computational domain. 
It is realized with a free slip condition for the velocity, a zero-gradient condition for pressure and a constant contact angle $\theta = \SI{90}{\degree}$.
To ensure the validity of this boundary condition, selected cases were simulated with the full geometry, confirming that the filling behavior is not noticeably influenced by the symmetry plane.
For brevity, these results are not shown here.
A constant velocity block profile is given at the inlet, and the outlet is defined as a zero-pressure boundary.
At the wall boundaries, a partial slip condition is used for the velocity, where the slip length is arbitrarily chosen as $\lambda = \SI{0.02}{mm}$.
The unstructured mesh of the domain is created with snappyHexMesh and described in~\cref{subsec:mesh}.
The time step for all cases is fixed to satisfy the capillary time step restriction 
\begin{equation} 
    \Delta t < \sqrt{\dfrac{(\rho^- + \rho^+) (\Delta x)^{3} }{4 \pi \sigma} },
    \label{eq:dt_capillary}
\end{equation}
where $\rho^\pm$ are the densities of the two fluids and $\Delta x$ is the mesh resolution, as discussed in~\cite{denner_numerical_2015}.
In the considered regimes, the capillary time step criterion is stricter than the CFL criterion.
A constant boundary condition for the contact angle is prescribed at the walls, the contact angles $\theta$ are chosen corresponding to the respective apparent contact angles measured in the experiments (\cref{subsubsec:03_dynCA_results},~\cref{fig:exp_dynCAfit}).
The contact angles are summarized in tab.~\ref{tab:sim_params}.
The dynamic contact angle $\theta_{\mathrm{cover}}$ between the fluid and the Zeonex cover plate, as introduced in~\cref{subsubsec:exp_materials}, cannot directly be measured with the used optical backlight setup, where only the interface in $xy$-plane can be analyzed.
For water and Tween, the equilibrium contact angle with Zeonex measured with standing droplet experiments is used instead.
For water, it equals $\theta = (94 \pm 5)\SI{}{\degree}$ and is in agreement with literature data~\cite{ganser_practical_2018, nagy_capillary_2022, albert_robust_2019}.
For Tween, the measured static contact angle is $\theta = (85 \pm 5)\SI{}{\degree}$.
For Novec, which is perfectly wetting, the static contact angle equals zero, therefore the dynamic contact angle is assumed to depend on the velocity in the same way as for the PMMA material.
The resulting values of $\theta_{\mathrm{cover}}$ are also summarized for all simulations in tab.~\ref{tab:sim_params}.
The results presented in~\cref{subsubsec:exp_filling_regimes} show that these assumptions yield a good match for the experimental results.
\begin{table}[]
\centering
\color{black}
\footnotesize
\caption{Contact angles and geometrical parameters for the simulation cases. The contact angles are given for the channel walls ($\theta$) and for the cover plate ($\theta_{\mathrm{cover}}$). The inflow length $L_{\mathrm{inflow}}$ and outflow length $L_{\mathrm{inflow}}$ are adapted to the capillary numbers and contact angles.}
\label{tab:sim_params}
\begin{tabular}{@{}llcccc@{}}
\toprule
Fluid & $Ca$ & $\theta \ (\SI{}{\degree})$  & $\theta_{\mathrm{cover}} \ (\SI{}{\degree})$  & $L_{\mathrm{inflow}}$ (mm) & $L_{\mathrm{outflow}}$ (mm) \\ \midrule
Water & $\SI{1.37e-4}{}$      & 112                               & 94                           & 1                  & 1                   \\
      & $\SI{1.37e-5}{}$      & 101                               & 94                           & 2                  & 1                   \\
Tween & $\SI{2.85e-4}{}$      & 102                               & 85                           & 1                  & 1                   \\
      & $\SI{2.85e-5}{}$      & 97                                & 85                           & 2                  & 1                   \\
Novec & $\SI{7.67e-3}{}$      & 45                                & 45                           & 2                  & 1                   \\
      & $\SI{7.67e-4}{}$      & 28                                & 28                           & 2                  & 8                   \\ \bottomrule
\end{tabular}
\end{table}
%
\section{Results and Discussion}
\label{sec:results}
\subsection{Dynamic Contact Angles}
\label{subsubsec:03_dynCA_results}
\subsubsection{Experimentally Acquired Contact Angles}\label{subsec:results_ca_exp}
\begin{figure}
    \centering
    \includegraphics[width=0.45\textwidth]{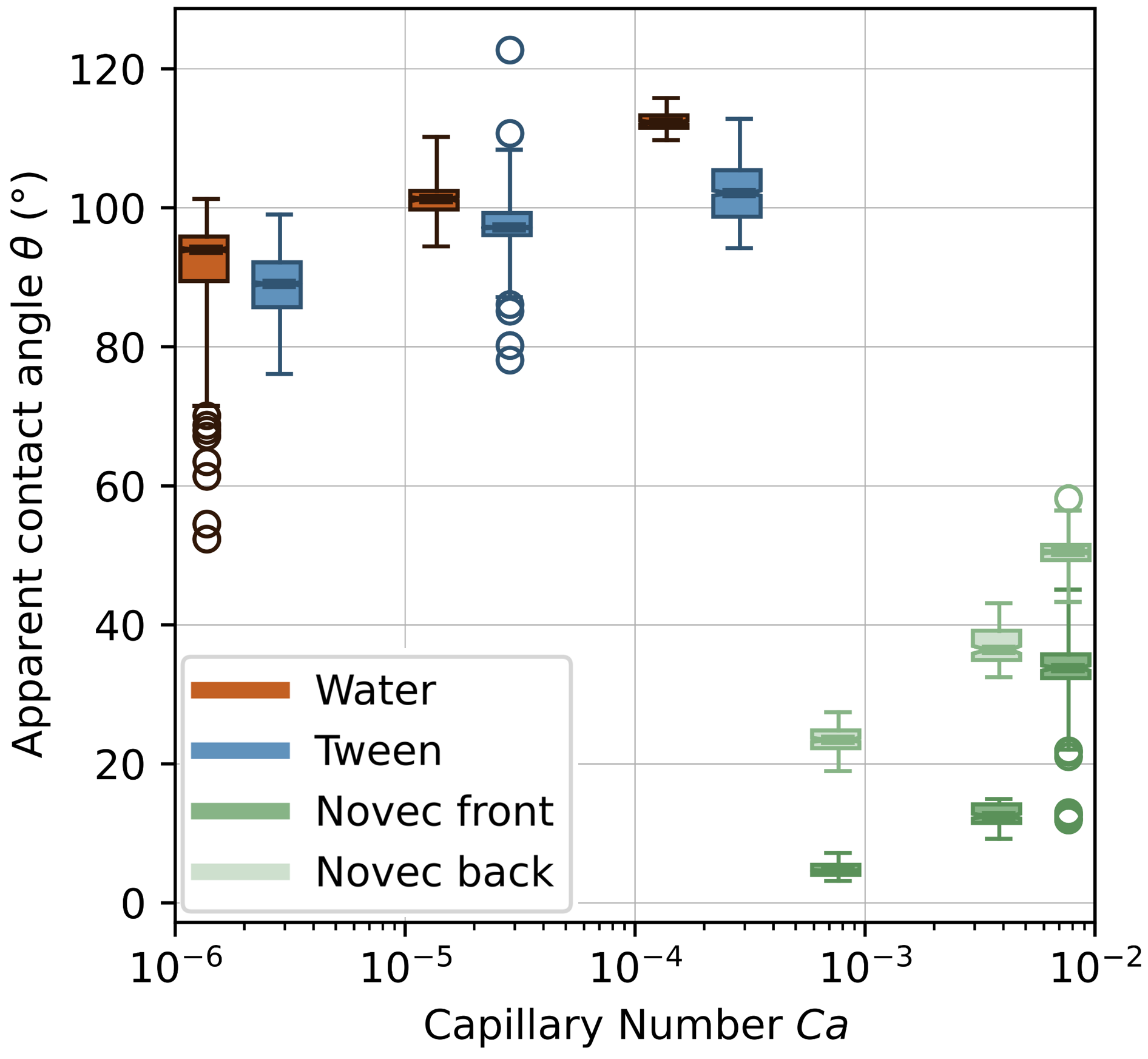}
    \caption{Measured apparent contact angles for the considered fluids. The horizontal lines show the median, the boxes represent the middle $\SI{50}{\percent}$ of the data. The whiskers extend from the box to the respective farthest data point lying within $3\times$ the height of the box, the circles represent outliers.}
    \label{fig:exp_dynCAfit}
\end{figure}
In~\cref{fig:exp_dynCAfit} the dependency of the apparent contact angles on the given capillary numbers is shown as a box plot.
\textcolor{black}{The values of the contact angles result directly from the automated data processing workflow described in~\cref{subsec:exp_data_processing}.
Each box plot includes the contact angle measurements at the channel top ($y=W$) and bottom ($y=0$) for the specified capillary number, as shown in~\cref{fig:exp_data_workflow}.
Due to the axial symmetry of the channel, the top and bottom contact angle measurements are grouped together.
Three repetitions of each experiment are used to achieve a robust dataset.}
A strong dependency of the contact angle on the capillary number is visible for all fluids, with the median contact angle increasing by more than ten degrees in the range of the given capillary numbers. 
The inter-quartile box representing the middle $\SI{50}{\percent}$ of the data spans over less than seven degrees, indicating that the medians represent the data sufficiently well.
For water and Tween, large outliers are visible in cases with apparent contact angles slightly above~$\theta = \SI{90}{\degree}$.
In this contact angle range, the interface appears very thin in the images, which reduces the detection accuracy.
For Novec, outliers occur at the highest capillary number, especially at the interface front, which can be attributed to single images in which small errors in the clustering step yield large approximation errors of the interface shape.
In general, the overall number of outliers is deemed acceptable for such a large dataset.
It should be noted that the local velocity in the vicinity of the wall can be expected to be slightly lower than the given mean velocity $U$ due to wall friction.
This influence is regarded as minor and subsequently neglected in this study.
\subsubsection{Contact Angle Calibration for Simulation}\label{subsec:ca_calibration}
\begin{figure*}
    \centering
    \includegraphics[width=0.9\textwidth]{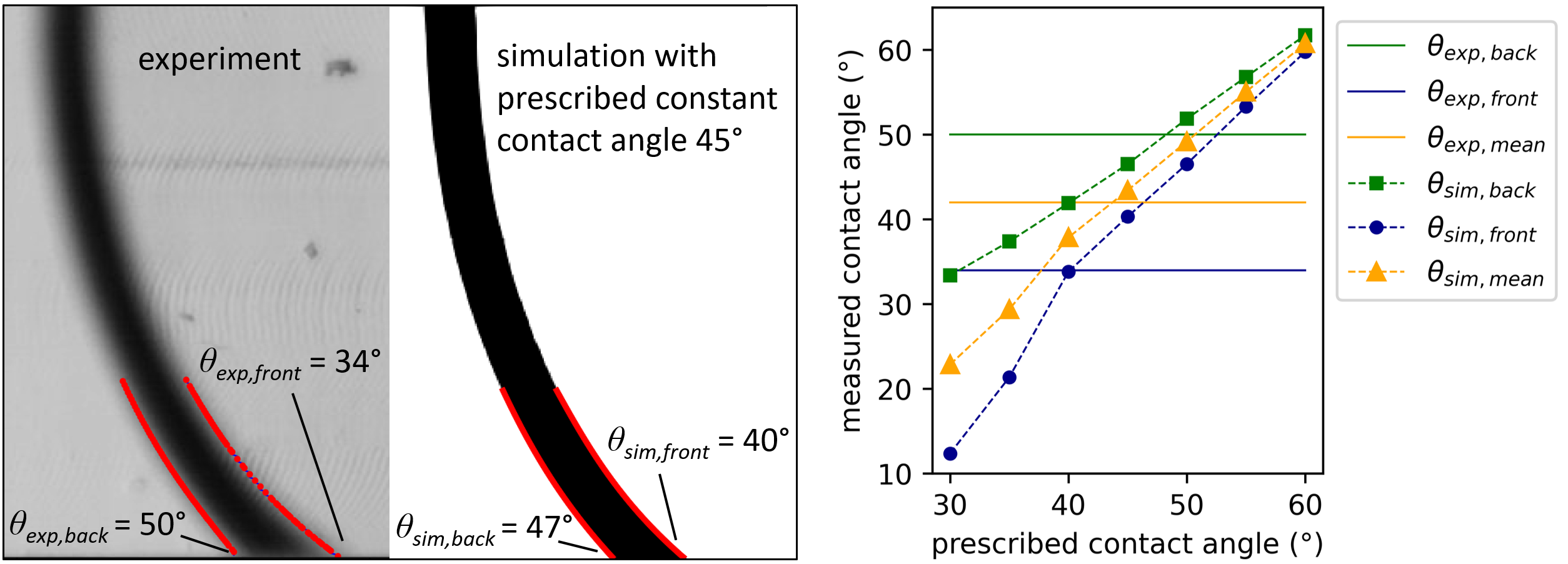}
    \caption{\textcolor{black}{Contact angle calibration for Novec at $Ca = \SI{7.67e-3}{}$.
    Left: comparison between contact angles in experiment and simulation. 
    In the simulation, a contact angle of $\SI{45}{\degree}$ is prescribed, resulting in a lower measured contact angle at the front and a higher measured contact angle at the back.
    Right: results of the full calibration study. The simulations were performed with different prescribed contact angles between $\SI{30}{\degree}$ and $\SI{60}{\degree}$, and the resulting measured contact angles are shown for the interface front and back, along with an averaged value. For comparison, the experimentally measured contact angles are given.}}
    \label{fig:ca_calibration_highCa}
\end{figure*}
\begin{figure*}
    \centering
    \includegraphics[width=\textwidth]{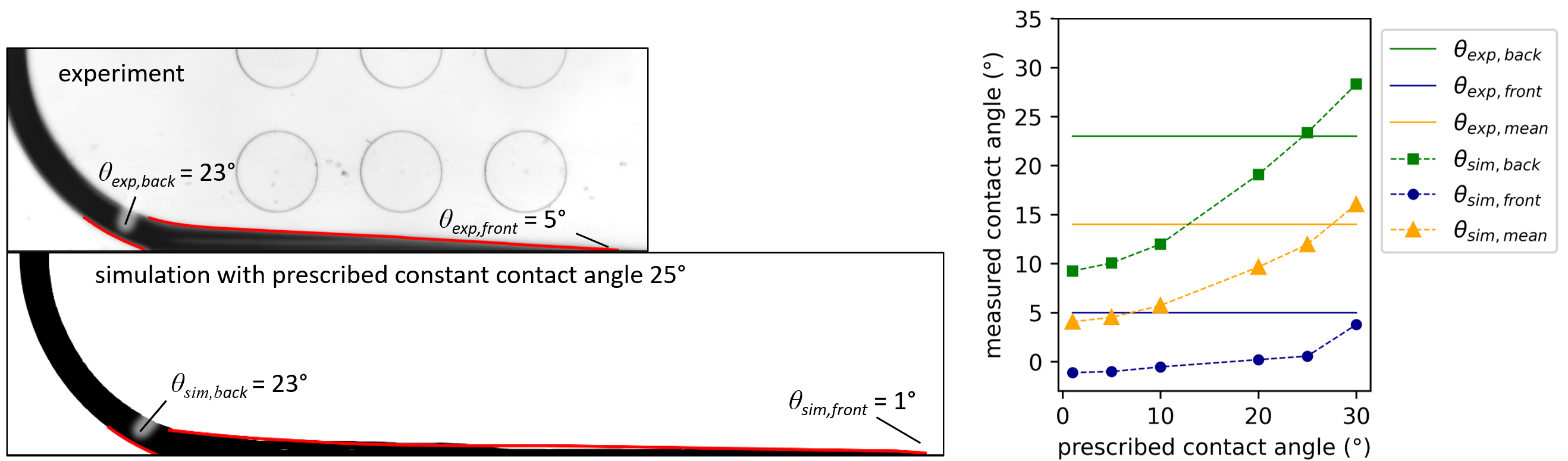}
    \caption{\textcolor{black}{Contact angle calibration for Novec at $Ca = \SI{7.67e-4}{}$. 
    Left: comparison between contact angles in experiment and simulation. 
    In the simulation, a contact angle of $\SI{25}{\degree}$ is prescribed, resulting in a significantly lower measured contact angle at the front and a slightly lower measured contact angle at the back.
    Right: results of the full calibration study. The simulations were performed with different prescribed contact angles between $\SI{1}{\degree}$ and $\SI{30}{\degree}$, and the resulting measured contact angles are shown for the interface front and back, along with an averaged value. For comparison, the experimentally measured contact angles are given.}}
    \label{fig:ca_calibration_lowCa}
\end{figure*}
For water and Tween, the medians presented in~\cref{fig:exp_dynCAfit} are used as constant contact angle boundary conditions in the simulations.
Since the pixel size of the images~${x_p = \SI{0.003}{mm}}$ is only one order of magnitude below the computational grid size $\Delta x=\SI{0.033}{\milli\meter}$, it is assumed that the measured apparent contact angles are suitable for this use.
\par
For Novec, the interface exhibits different apparent contact angles at the front and the back, respectively.
\textcolor{black}{This is shown in the left subfigure of~\cref{fig:ca_calibration_highCa} for the capillary number $Ca = \SI{7.67e-3}{}$.
The front and back contact angle are $\theta_{exp, front}=\SI{34}{\degree}$ and $\theta_{exp, back}=\SI{50}{\degree}$, respectively.
In the simulation, however, only a single contact angle can be prescribed.
This raises the question of which contact angle should be chosen in the simulation to represent the experimental data.
To investigate this, simulations of a channel flow, excluding the cavities, are carried out with different contact angles to find the best match for the experiments.}
To achieve an appropriate comparison, the simulation data needs to be post-processed with the same algorithm as the experimental images, cf.~\cref{subsec:porproc_novec}. 
Therefore, to get a matching visual representation of simulation and experiment, the experimental view is reproduced in the simulations using a 2D projection of the reconstructed interface in ParaView~\cite{Ahrens2005ParaViewAE}.
\textcolor{black}{This is shown in the center subfigure of~\cref{fig:ca_calibration_highCa} for a simulation with prescribed contact angle $\theta=\SI{40}{\degree}$.
It is apparent that the measured contact angles $\theta_{sim, front}=\SI{40}{\degree}$ and $\theta_{sim, back}=\SI{47}{\degree}$ do not match the prescribed contact angle.}
Apart from the uncertainties induced by the 2D projection and the image processing workflow, which are assumed to be in the order of up to five degrees, a further source of uncertainty is the implementation of the contact angle model in the simulation framework.
\textcolor{black}{As recently discussed in~\cite{asghar2023numerical}, the contact angle is indirectly considered in the interface reconstruction by employing a ghost point method. 
In this method, the contact angle boundary condition influences the angle of the discrete interface, but the angle is not directly enforced.
A detailed description can be found in~\cite{scheufler_accurate_2019}.
This implementation gives rise to inaccuracies, especially for very high or very low contact angles.}
\textcolor{black}{
To determine which contact angle needs to be prescribed in the simulation to match the experimental results best, a line-search parameter study is carried out.
Simulations with contact angles $\theta=30...60 \SI{}{\degree}$ are conducted, and the resulting front and back contact angles are presented in the plot in~\cref{fig:ca_calibration_highCa}. 
As a comparison, the experimental contact angles are shown as horizontal lines.
To identify the simulation that matches the experimental data best, the average of the front and back contact angles, $\theta_{sim, mean}$ and $\theta_{exp, mean}$ are compared.
It is found that the simulation with the prescribed contact angle $\theta=\SI{45}{\degree}$ represents the experimental results best. }
\par
\textcolor{black}{For $Ca = \SI{7.67e-4}{}$, the deviation between the measured front and back contact angle is large, both in the experiments and the simulations, as shown in~\cref{fig:ca_calibration_lowCa}}.
This can be explained by the long rivulets exhibited at the channel edges, which pose a challenge not only for the interface reconstruction in the simulation but also for the image processing algorithm.
Especially in the experimental data, the tips of thin rivulets are difficult to distinguish from the wall in the grayscale images due to limited resolution.
Moreover, the rivulet shape depends strongly on the edge geometry, and while a perfect $\SI{90}{\degree}$ wall angle is used in the simulations, the experimental geometry might deviate slightly, adding another source of uncertainty.
Generally, the formation of rivulets during forced wetting in rectangular channels is an active research topic in itself (cf.~\cite{gerlach_rivulets_2021, thammanna_gurumurthy_forced_2022}) and an in-depth analysis of their length and shape is beyond the scope of this work.
Instead, the focus here is on establishing a suitable contact angle for the simulations, which is in the range of $\theta=25...30\SI{}{\degree}$ according to the graph in\textcolor{black}{~\cref{fig:ca_calibration_lowCa}.
The visual comparison of the experiment and the simulation with $\theta=\SI{25}{\degree}$ (\cref{fig:ca_calibration_lowCa} left) shows a good qualitative match of the interface shape with the exception of the rivulet length.}
For the following sections, the contact angle $\theta=\SI{28}{\degree}$ is chosen as a compromise between the two most suitable contact angles $\theta=\SI{25}{\degree}$ and $\theta=\SI{30}{\degree}$.
\par 
To test the robustness of the presented results, the dependency of the measured contact angles on mesh resolution, artificial viscosity, and slip length is shortly discussed in~\ref{app:ca_caliration}.
\subsection{Mesh Refinement Study}
\label{subsec:mesh}
\begin{figure*}
    \centering
    \includegraphics[width=\textwidth]{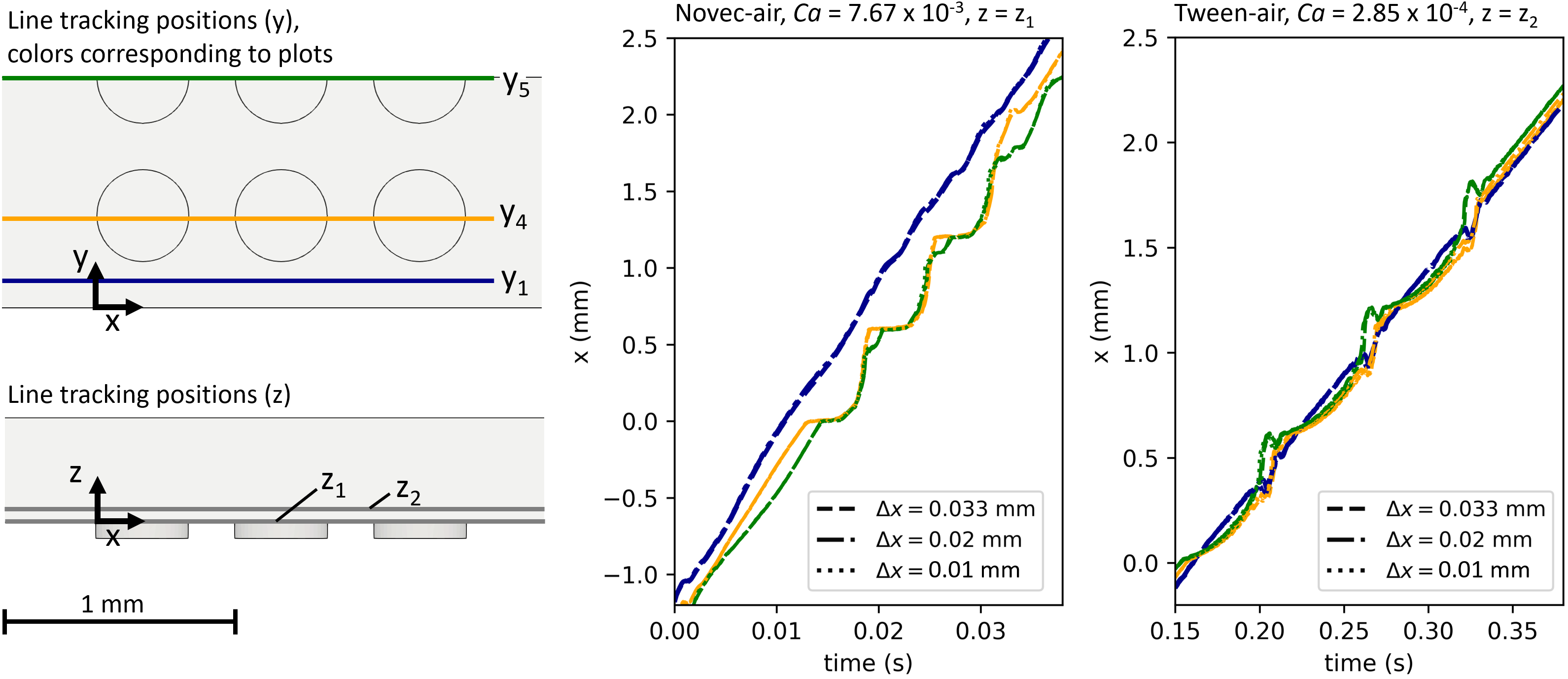}
    \caption{\textcolor{black}{Setup and results for the mesh study using interface tracking. 
    Left: Schematic of the positions along which line tracking is performed in the simulations.
    The three different $y$ positions are indicated by colors corresponding to the line colors in the plots. 
    Center and right: Line tracking results for two different cases.
    For Novec-air at ${Ca = \SI{7.67e-4}{}}$, the results are given at $z=z_1$.
    For Tween-air at~${Ca = \SI{2.85e-4}{}}$, the results are given at $z=z_2$.
    The line styles represent the mesh resolutions. 
    }}
    \label{fig:sim_meshstudy}
\end{figure*}
To evaluate the dependency of the simulation results with regard to increasing mesh resolution, two cases are considered which represent different filling regimes~(cf.~\cref{subsubsec:exp_filling_regimes}).
Three different mesh resolutions are considered, with cell lengths $\Delta x=(0.033,$, $0.02$, $0.01)\, \SI{}{\milli\meter}$.
As a means of comparison, the interface movement at different positions is used.
\textcolor{black}{The interface is tracked along three horizontal lines, ${y_{1,4,5} = (0.0001, 0.1, 10) \, \SI{}{mm}}$, which are shown as colored lines in the sketch on the left side in~\cref{fig:sim_meshstudy}.
The results are given in the center and right subfigures of~\cref{fig:sim_meshstudy}, where the line colors indicate the corresponding $y$ positions in the sketch.}
\par
In the \textcolor{black}{center} plot in~\cref{fig:sim_meshstudy}, the result for the case with the fluid combination Novec-air at ${Ca = \SI{7.67e-4}{}}$ is displayed.
In this case, the cavities are filled, as discussed later in~\cref{subsubsec:filled}.
The interface is tracked directly above the cavities at ${z = z_1 = \SI{0.001}{mm}}$ across \textcolor{black}{the three $y$ positions.
The different meshes are indicated with different line styles.}
It can be seen that the interface movements at the different $y$~positions differ significantly, but the difference between the resolutions is indistinguishable in comparison.
The same holds for the second case (Tween-air at~${Ca = \SI{2.85e-4}{}}$) which is shown in the right plot.
Here, the cavities are not filled, as discussed later in~\cref{subsubsec:notfilled}.
In this scenario, the interface is tracked above the cavities at ${z = z_2 = \SI{0.05}{mm}}$ to exclude the impact of the interface pieces separating at the top of the cavities. 
The interface movement across several $y$~positions does not differ strongly, and the changes in the mesh resolution do not influence the progression significantly.
As a result of these studies, the mesh resolution~${\Delta x = \SI{0.033}{\milli\meter}}$ is chosen for all following simulations as a compromise between computational resources and a well-resolved interface.
\subsection{Cavity Filling Regimes}
\label{subsubsec:exp_filling_regimes}
\begin{figure*}
    \centering
    \includegraphics[width=\textwidth]{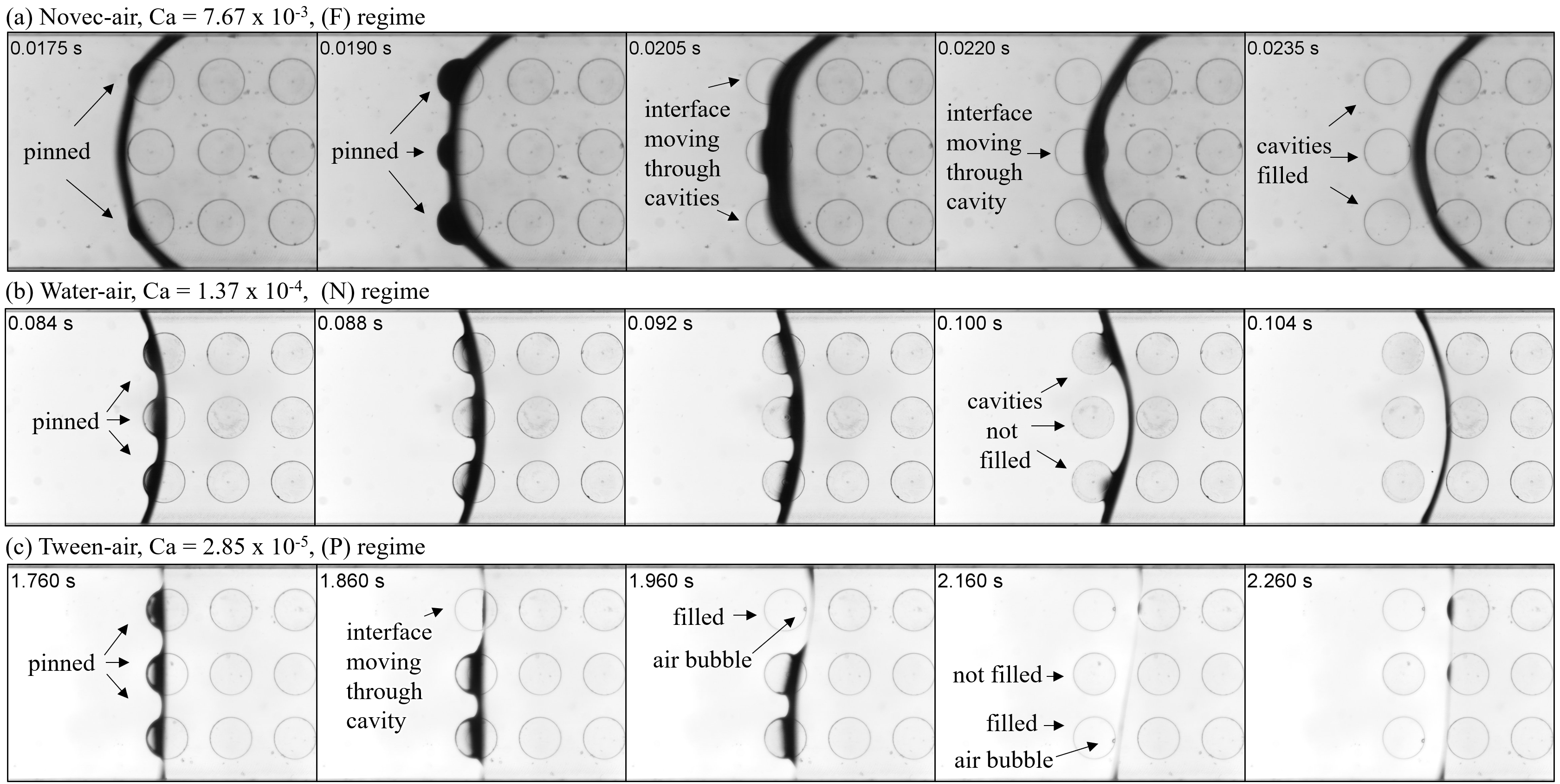}
    \caption{Examples of the interface movement with different fluids and \textcolor{black}{capillary numbers}.}
    \label{fig:exp_examples}
\end{figure*}
\subsubsection{Regime Distinction}
When the interface passes the cavities, three scenarios are possible.
Examples are depicted in~\cref{fig:exp_examples}, where the regimes are differentiated as follows:
\begin{itemize}
    \item[--] Filled (F),~\cref{fig:exp_examples}a: 
    During the filling process, the interface first pins at the cavity edges, and subsequently moves through the cavities.
    All nine cavities are filled with the fluid, in some cases leaving tiny air bubbles at the cavity bottoms. 
    \item[--] Not Filled (N),~\cref{fig:exp_examples}b: 
    Pinning is observed at the cavity edges, and the interface moves over the cavities, leaving behind air cushions.
    No cavity is fully filled. 
    \item[--] Partially Filled (P),~\cref{fig:exp_examples}c: Some of the cavities are fully filled, while air cushions or large bubbles remain in some other cavities.
\end{itemize}
To determine whether the cavity array is filled, partially filled, or not filled, the image sequence is interpreted with regard to the movement of the interface and the grayscale values of the cavities after the interface has passed.
All experiments are performed three times to ensure reproducibility, and the filling regimes are confirmed in the repeated measurements.
\begin{figure*}
    \centering
    \includegraphics[width=\textwidth]{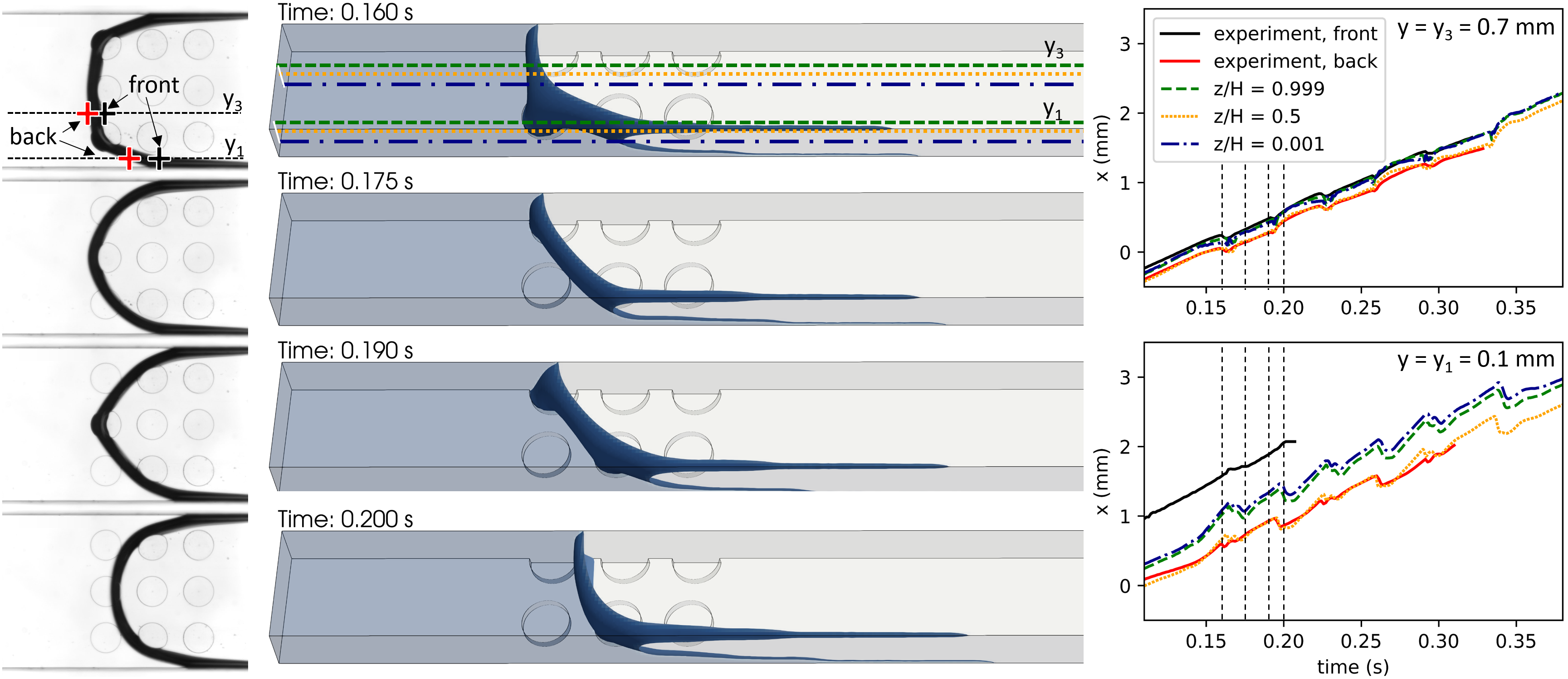}
    \caption{Interface tracking results for (F) regime, fluid pairing Novec-air, $Ca = \SI{7.67e-4}{}$. Left: experimental view, center: simulation results including positions of the tracking lines, right: interface tracking plot at two $y$ positions\textcolor{black}{: $y=y_3$ (upper subfigure), $y=y_1$ (lower subfigure).}}
    \label{fig:sim_tracking_fill}
\end{figure*}
%
\subsubsection{(F) Regime - Complete Cavity Filling} 
\label{subsubsec:filled}
In \cref{fig:exp_examples}a the (F) regime is depicted for the experiment with Novec at $Ca = \SI{7.67e-3}{}$.
It is characterized by the interface first pinning at the cavity edges before traversing them.
A more close-up look at the filling is given in~\cref{fig:sim_tracking_fill} for the experiment with Novec~at $Ca = \SI{7.67e-4}{}$.
In the experimental images (\cref{fig:sim_tracking_fill} left) the interface appears thick, and rivulets are visible at the walls due to the low contact angle.
Four time steps are shown which represent critical steps during the filling process.
First, the interface pins at the edges of the first cavity row ($t = \SI{0.16}{s}$). 
Next, within approximately $0.015$ seconds, the near-wall cavities are filled ($t = \SI{0.175}{s}$). 
Afterward, the interface pins strongly at the center cavity ($t = \SI{0.19}{s}$) and traverses it ($t = \SI{0.2}{s}$). 
This highly dynamic process is confirmed by the simulation results (\cref{fig:sim_tracking_fill} center).
The 3D view reveals that the interface is strongly curved across the $y$ axis, therefore appearing very thick in the top-down view.
\par 
For a quantitative analysis of the filling process, interface tracking plots are used (\cref{fig:sim_tracking_fill} right).
The experimental results are given in terms of the interface front and back pixels that are identified as discussed in~\cref{subsec:porproc_novec}.
Two positions are considered for the interface tracking: $y=y_1=\SI{0.7}{mm}$ between two cavity rows, and $y=y_2=\SI{0.1}{mm}$ between the lower cavity row and the wall.
The results are given as solid lines in the graphs, limited by the size of the field of view spanning two millimeters.
It is apparent that at $y_1$, the front and back are close to each other, whereas at $y_2$, they are approximately one millimeter apart, which represents the long rivulets at the channel edges.
Furthermore, kinks can be seen in the interface progression at both positions.
The times at which these kinks appear match the pinning and filling steps visualized in \cref{fig:sim_tracking_fill} left.
To compare the numerical results to the experiment, the same $y$ positions are considered for interface tracking in the simulation.
Additionally, the 3D  results allow the tracking along different heights $z$, which are represented by different colors in the graph.
At $y_1$, the simulation results show a very close match to the experiment, including the kinks.
At $y_2$, a discrepancy is visible between simulation and experiment.
The simulation result at $z/H=\SI{0.5}{}$ matches the back of the experimental interface.
In the 3D view, this corresponds to the center of the interface which lags behind the top and bottom due to the low contact angle.
With this reasoning, the experimental interface front is expected to match the simulation at $z/H=\SI{0.001}{}$ and $z/H=\SI{0.999}{}$.
However, while the trend in the lines is correct, a correct match is not achieved.
This can be attributed to the difficulties in correctly representing rivulet length and shape, as discussed in~\cref{subsec:ca_calibration}.
Due to the good match of the other characteristics and the filling state, which remains the main focus of this study, the results are considered satisfactory.
%
\subsubsection{(N) Regime - No Cavity Filling}  
\label{subsubsec:notfilled}
In the (N) regime (\cref{fig:exp_examples}b), the interface pins at the cavity edges, and moves over the cavities.
No snapping movement through the cavities is observed, and air cushions remain in the cavities, covered by the fluid.
It should be noted that the interface covering the cavities is parallel to the camera plane, and is therefore detectable only due to a minor decrease in brightness.
The process of pinning and detachment generally occurs symmetrically to the channel center line, but in some repetitions, the interface at one wall proceeds the movement of the opposite side, which can be caused by minor differences in roughness or contamination.
A further observation is that in the experiments with Tween, the air cushions tend to change shape to a more spherical bubble over time.
The details of this process cannot be investigated within the scope of this work, however, a possible explanation is a transient behavior of the surfactant distribution along the interface.
\begin{figure*}
    \centering
    \includegraphics[width=\textwidth]{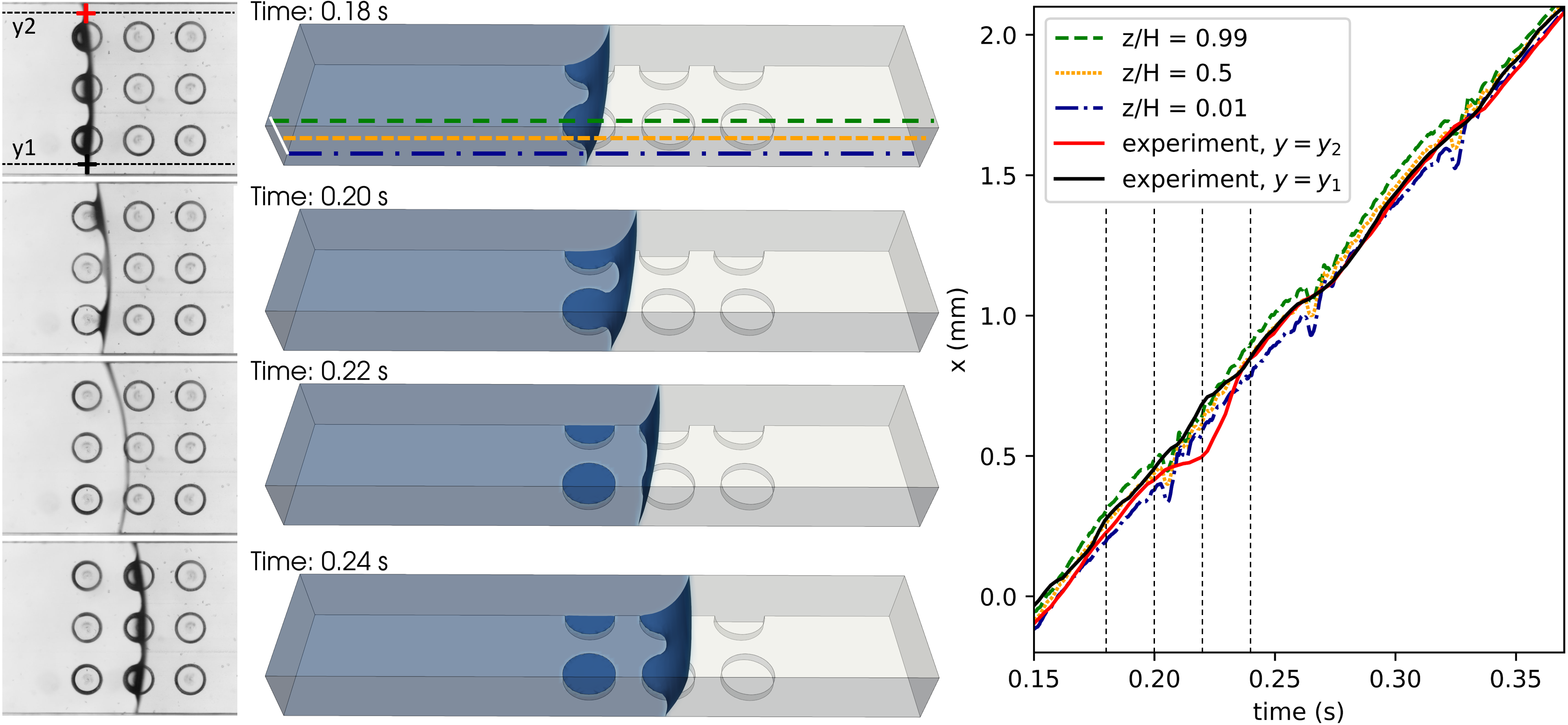}
    \caption{\textcolor{black}{Interface tracking results for (N) regime, fluid pairing Tween-air, $Ca = \SI{2.85e-4}{}$. Left: experimental view, center: simulation results including positions of the tracking lines, right: interface tracking plot.}}
    \label{fig:sim_tracking_notfill}
\end{figure*}
\par
Fig.~\ref{fig:sim_tracking_notfill} presents the flow of Tween \ at $Ca = \SI{2.85e-4}{}$, along with the comparison to simulation results.
The visual comparison of the interface movement over the first cavity row is shown at the left and center of~\cref{fig:sim_tracking_notfill}.
The pinning of the interface ($t=\SI{0.18}{s}$) is succeeded by the detachment at the center cavity ($t=\SI{0.2}{s}$) and at the wall-adjacent cavities~($t=\SI{0.22}{s}$).
Subsequently, the process is repeated at the second cavity row ($t=\SI{0.24}{s}$).
\par
The quantitative comparison of the interface movement is shown on the right side of~\cref{fig:sim_tracking_notfill}.
The experimental results are displayed for the top and bottom interface movement, corresponding to the positions given in the top left picture of~\cref{fig:sim_tracking_notfill}.
It can be observed that the interface movement is relatively uniform with the exception of the time span $\SI{0.2}{s} \leq t \leq \SI{0.24}{s}$. As seen in the visual depictions, this time span corresponds to the interface moving over the first cavity row.
In the experiment, the interface detaches slightly earlier at the bottom cavity than the top cavity, resulting in an asymmetric shape, visible at $t = \SI{0.22}{s}$.
\textcolor{black}{This causes the deviation between top and bottom curves in the interface tracking plot in~\cref{fig:sim_tracking_notfill}.}
However, for the second and third cavity rows, the interface tracking is more uniform, only showing slight waves when the cavities are passed, signifying a symmetrical interface shape.
\par In the simulations, the general movement is confirmed.
\textcolor{black}{Kinks in the interface progression are visible at the times when the cavity rows are passed, showing an oscillation when the air cushions are created and the interface breaks up into separate parts: the progressing interface and the interface sections covering the air cushions in the cavities.}
These oscillations occur slightly earlier than in the experimental curve, signifying a minor difference in the timing of the interface separation.
\textcolor{black}{Furthermore, due to the symmetric simulation setup, the asymmetric interface movement of the experiments at $\SI{0.2}{s} \leq t \leq \SI{0.24}{s}$ cannot be reproduced in the simulations.}
Besides this, the experimental results are very well matched.
The very low deviation between the line tracking results at different heights confirm a small interface deformation compared to the (F) case, which is confirmed by the thin appearance of the interface in the experimental image. 
%
\subsubsection{(P) Regime - Partial Cavity Filling}
\label{subsubsec:partial}
In the (P) regime (\cref{fig:exp_examples}c) the exact movement of the interface and the number of filled cavities are not directly reproducible between repeated experiments.
The filling process in this regime depends heavily on local effects such as pinning on small roughness edges.
In this regime, the simulation results deviate from the experimental data.
The local deviations from the idealized sharp-edge geometry cannot be represented in the simulations, and the simulation yields a (N) regime.
One further aspect to consider is the effect of the dynamic contact angle. 
In the simulation, the velocity dependence of the contact angle is not captured, which is also a possible explanation for the discrepancies.
To put the (P) regime into context, it should be noted that partial filling is not desired in LoC applications. 
The appearance of even a few small air bubbles in large-scale microcavity arrays can compromise the accuracy of the biochemical processes.
Therefore, the tendency of the simulation towards predicting the (N) regime can be considered as a conservative prediction with regard to the filling state.
This is a positive aspect for future exploratory studies using simulations.
%
\begin{figure*}
    \centering
    \includegraphics[width=0.99\textwidth]{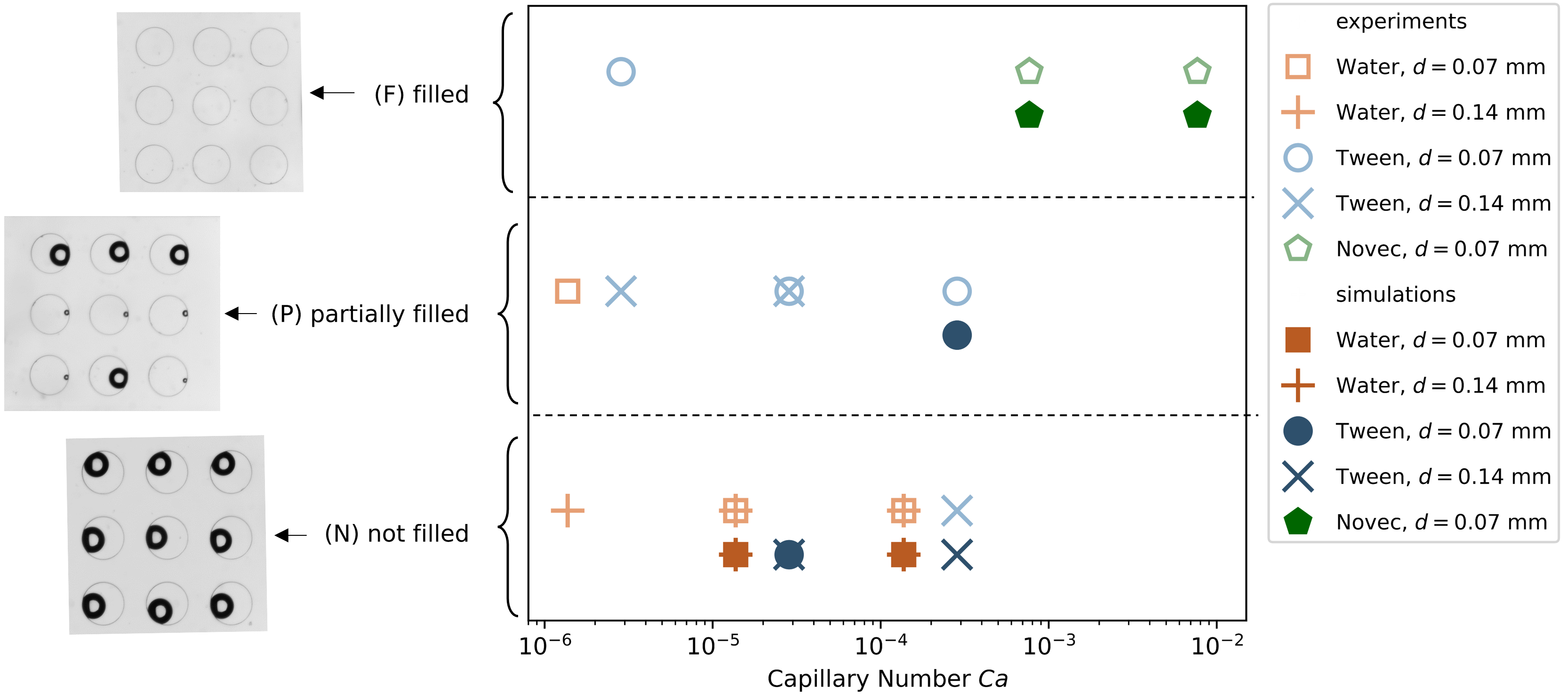}
    \caption{\textcolor{black}{Filling regime map distinguishing three different regimes. Experimental data is shown with light colors, simulation results are displayed with dark colors.}}
    \label{fig:exp_fillingRegimes}
\end{figure*}
\subsubsection{Regime Map}
\label{subsubsec:regime_map}
To summarize the results,~\cref{fig:exp_fillingRegimes} displays the filling regimes acquired in all experiments.
\textcolor{black}{The filling states are presented for different fluids, capillary numbers, and, for Tween and water, two different cavity depths.
The experimental results are displayed with light-colored symbols.
Additionally, the simulation results, which are explained further below, are represented with dark colors.}
The experiments with Novec result in the (F) regime in all cases.
It can be assumed that for lower capillary numbers, the (F) regime is also achieved since capillary forces will dominate even stronger and the contact angle will be reduced even further.
In the experiments with Tween, different regimes are observed depending on the capillary number and the cavity depth.
The experiments show that increasing capillary numbers result in an overall lower chance of filling.
This can be attributed to the decrease in importance of capillary force, and to the increasing contact angles with increasing capillary number, as seen in~\cref{fig:exp_dynCAfit}.
Furthermore, the deep cavities ($d=\SI{0.14}{mm}$) are less likely to be filled than the shallow cavities ($d=\SI{0.07}{mm}$).
This result is intuitive and matches the findings in~\cite{padmanabhan_enhanced_2020}.
The experiments with water yield the (N) regime in all but one scenario.
The high contact angle prevents the filling, and only the lowest capillary number results in a partial filling in the shallow cavities. 
\par
In addition to the experimental data,~\cref{fig:exp_fillingRegimes} gives an overview of the simulation results\textcolor{black}{, which are represented with dark colors.}
Not all capillary numbers are considered in the simulations since simulations with low capillary numbers ($Ca < \SI{e-5}{}$) show poor stability due to high parasitic currents.
In the considered range of capillary numbers, the simulations match the experimentally acquired regimes very well for the (F) and (N) cases, as discussed in~\cref{subsubsec:filled} and~\cref{subsubsec:notfilled}.
The (P) regime is not achieved in any of the simulations for the reasons discussed in~\cref{subsubsec:partial}.
In general, the agreement between simulations and experiments is very satisfactory, indicating that VoF simulation is a suitable tool to predict cavity filling states.
Moreover, the simulations provide 3D insights into the interface shape and movement, which cannot be achieved in the top-down view of the experiments, and therefore help to explain the experimentally achieved results.
%
\section{Summary and Conclusion}
\label{sec:conclusion}
Understanding the mechanisms of a fluid interface traversing over microcavities is vital for improving the efficiency of microcavity filling for LoC applications.
This work presents a close-up experimental investigation of filling regimes of a $3\times3$ microcavity array with various fluids and capillary numbers.
The apparent dynamic contact angles and interface movements are acquired with an automated image processing workflow, ensuring reproducible results, and providing calibration data for simulations.
 \textcolor{black}{
The close-up examination of the interface movement during various filling scenarios reveals distinct mechanisms across three regimes: filled, not filled, and partially filled. Filling is shown to be more likely at low capillary numbers. 
Lower contact angles increase the filling probability, while deeper cavities reduce it. 
To further explore the separate impacts of contact angle and capillary number, a dedicated study is recommended. 
In the present work, in which the fluid and the inflow velocity are varied, the effects of contact angle and capillary number cannot be separated, since the fluid properties and flow velocity impact both the capillary number and the contact angle. 
Future work could achieve a variation of contact angles at a constant capillary number by altering the sample material while maintaining a constant inflow velocity.}
\par 
 \textcolor{black}{In addition to the experimental results, it is shown that the simulation using the geometrical unstructured VoF method can predict the filling behavior for the filled and not filled regimes.}
The third regime, in which only a part of the cavities is filled, cannot be recreated by the simulations.
In this regime, the filling state depends heavily on minor geometry kinks, transient surfactant distribution, and contaminations, which cannot be represented in the simulations.
Since partial filling is not desired in LoC applications, the tendency of the simulation towards not filling the cavities is a conservative prediction.
\par
Overall, the fact that simulations can predict the filling state across a range of capillary numbers and contact angles is a promising finding.
The 3D insights into the interface shape acquired in the simulations complement the 2D experimental data.
In future work, such simulations can be used to perform shape optimizations and large parameter studies that would be very costly to perform experimentally.
Furthermore, additional effects such as species transport can be investigated in future work to shed more light on reactive multiphase flows occurring in LoC systems.
\section*{Acknowledgments}
The authors would like to thank Dr. Mathis Fricke from Technical University Darmstadt for the helpful and valuable discussions on contact line physics. 
\textcolor{black}{The authors express their gratitude to Reviewer 1 for providing a detailed review which has significantly helped improve the quatlity of the manuscript.}
\section*{CRediT authorship contribution statement}
\textbf{Luise Nagel:} Conceptualization, Methodology, Investigation, Software, Data curation, Validation, Visualization, Writing - original draft.
\textbf{Anja Lippert:} Conceptualization, Supervision, Project administration, Writing – review \& editing.
\textbf{Ronny Leonhardt:} Methodology, Resources, Writing – review \& editing.
\textbf{Tobias Tolle:} Supervision, Project administration, Writing – review \& editing.
\textbf{Huijie Zhang:} Writing – review \& editing.
\textbf{Tomislav Maric:} Conceptualization, Supervision, Methodology, Data curation, Writing – review \& editing.
\section*{Funding sources}
The last author acknowledges the funding by the German Research Foundation (DFG): 1 July 2020 - 30 June 2024 funded by the German Research Foundation (DFG) - Project-ID 265191195 - SFB 1194.
\appendix
\section{Accuracy of Polynomial Fit in the Interface Detection Method}
\label{app:rms}
\setcounter{figure}{0}    
\begin{figure*}
    \centering
    \includegraphics[width=0.7\textwidth]{{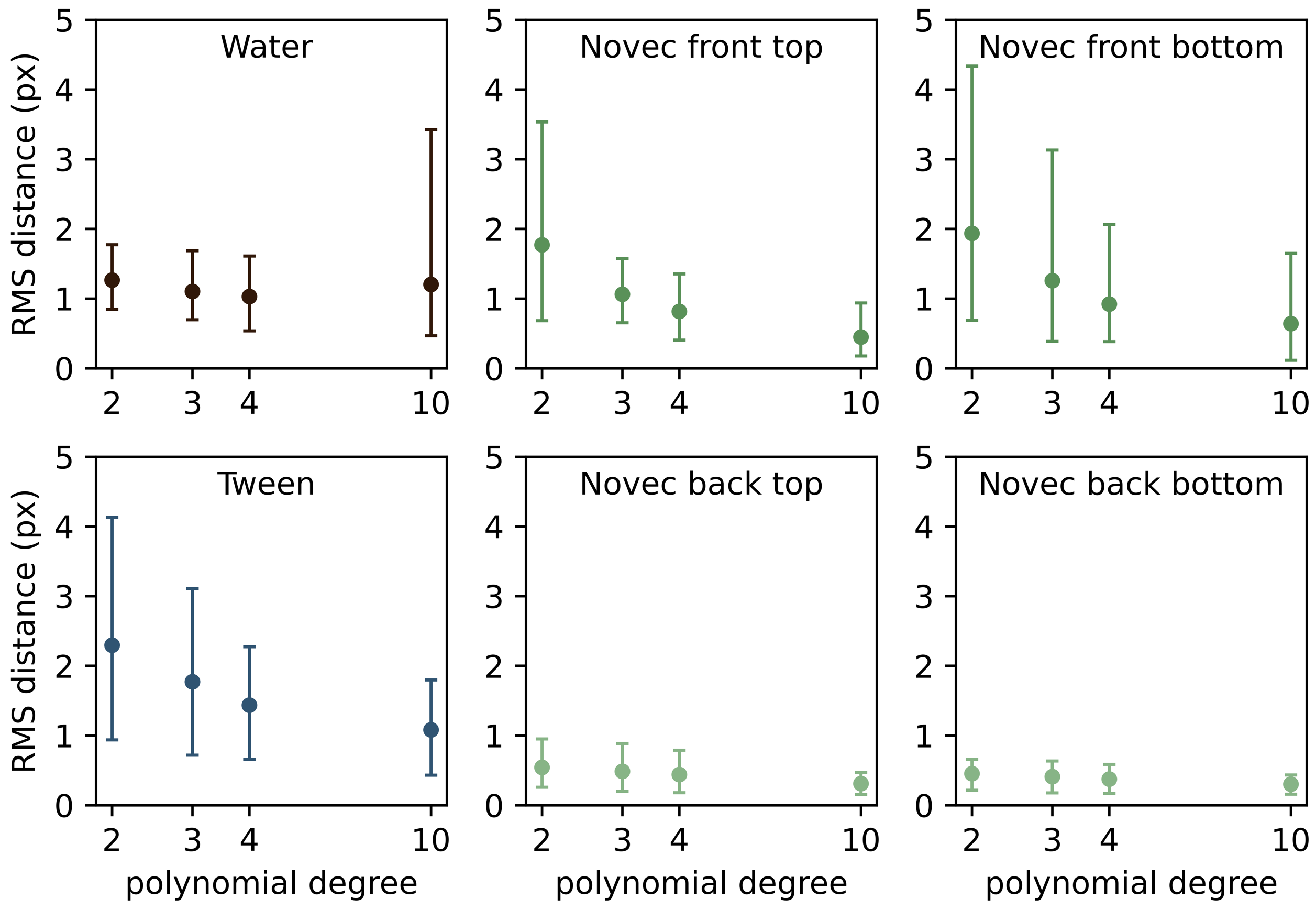}}
    \caption{Distance between sampled inlier points and polynomial fit in interface detection for all considered fluids.}
    \label{fig:rms_distance}
\end{figure*}
To test the accuracy of the polynomial fit applied to the interface shape, the root-mean-square distance $x_{RMS}$ between the inliers of the RANSAC method and the resulting polynomial is computed for all images of each case as
\begin{equation}
    x_{RMS} = \frac{1}{N_{img}} \sum_{N_{img}} \left[ \sqrt{\frac{1}{N_{in}} \sum_{N_{in}}\left((x_{in} - x_{fit})^2 \right)}\right],
\end{equation}
where the index $img$ refers to the images, $in$ refers to the inliers, and $x_{fit}$ indicates the points of the polynomial fitting curve.
To test the influence of the polynomial degree on the fit accuracy, four different polynomial degrees are compared.
Fig.~\ref{fig:rms_distance} shows the result for all considered fluids as error bar plots, in which the dots signify the average of $x_{RMS}$ and the bars signify the maximum and minimum values.
For Novec, the interface is fitted separately at different positions, therefore the respective results are given in separate plots.
It can be seen that the fit accuracy is within five pixels in all cases.
Since the influence of the polynomial degree is sufficiently small, the choice of a third-degree polynomial for all measurements is justified.
\section{Contact Angle Calibration Study}\label{app:ca_caliration}
\setcounter{figure}{0}    
\begin{figure*}
    \centering
    \includegraphics[width=\textwidth]{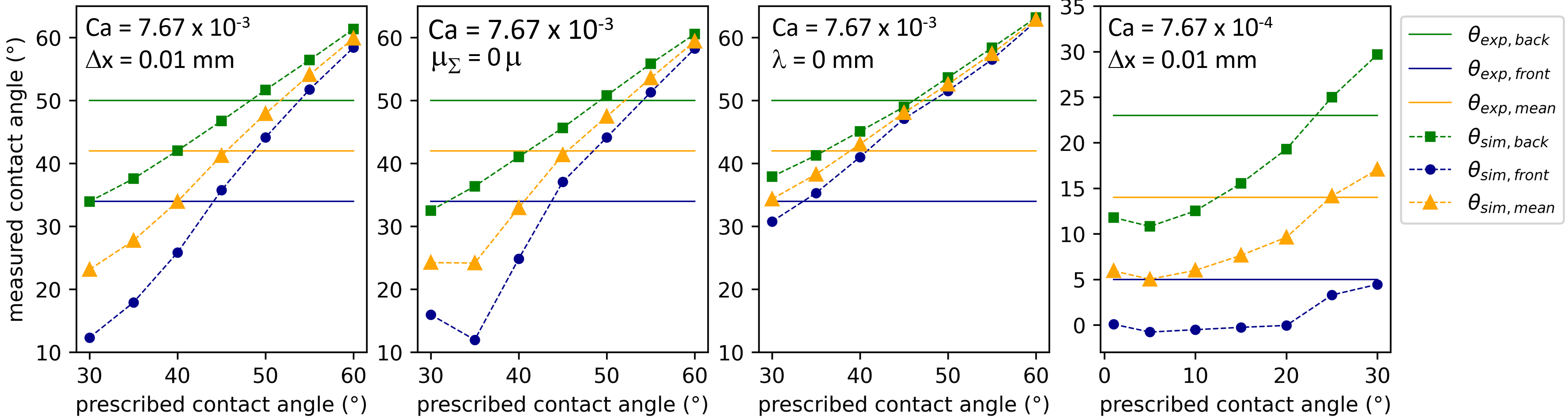}
    \caption{Contact angle calibration study with respect to different simulation parameters.}        
    \label{fig:ca_calibration_appendix}
\end{figure*}
To extend the results presented in~\cref{subsec:ca_calibration}, the dependency of the contact angle calibration on different parameters is shortly discussed here.
Fig.~\ref{fig:ca_calibration_appendix} shows the calibration study results for Novec at $Ca=\SI{7.67e-3}{}$ in three configurations: With increased mesh resolution ($\Delta x=\SI{0.01}{\milli \meter}$), without artificial viscosity ($\mu_\Sigma=\SI{0}{\mu}$), and with a no-slip boundary condition at the walls ($\lambda=\SI{0}{\milli \meter}$), respectively.
For the cases with  $\Delta x=\SI{0.01}{\milli \meter}$ and $\mu_\Sigma=0$, only slight deviations in the measured contact angles are visible compared to the graph in~\cref{fig:ca_calibration_highCa}.
These deviations of up to five degrees are in the order of magnitude of the general uncertainty of the method, and are therefore not considered significant.
In the case of $\lambda=\SI{0}{\milli \meter}$, the measured contact angles are closer to the prescribed contact angles and generally higher than for $\lambda=\SI{0.02}{\milli \meter}$ as used in~\cref{subsec:ca_calibration}. 
This shows that the slip length significantly influences the apparent contact angle in the simulation, and care should be taken in the calibration to account for the chosen slip length.
A further investigation of this effect is planned in future work.
However, the average apparent contact angle which matches the experimental value best, $\theta =\SI{40}{\degree}$, is still within five degrees of the originally established value $\theta =\SI{45}{\degree}$ presented in~\cref{subsec:ca_calibration}, and therefore acceptable in the scope of the present study.
\par
The rightmost figure in~\cref{fig:ca_calibration_appendix} presents an additional graph for Novec at $Ca=\SI{7.67e-4}{}$ with increased mesh resolution ($\Delta x=\SI{0.01}{\milli \meter}$). 
In comparison to~\cref{fig:ca_calibration_lowCa}, the result confirms the low mesh dependency of the measured contact angles previously discussed for $Ca=\SI{7.67e-3}{}$.





\end{document}